\documentclass{ws-ijmpcs}

\def\trimmarks{}

\begin{document}

\markboth{Masashi Wakamatsu}
{Are there infinitely many decompositions of the nucleon spin ?}

%
\catchline{}{}{}{}{}
%

\title{Are there infinitely many decompositions of the nucleon spin ?
}

\author{Masashi Wakamatsu}

\address{Department of Physics, Faculty of Science, \\ Osaka University,
Toyonaka, Osaka 560-0043, Japan}

\maketitle

\begin{history}
\received{Day Month Year}
\revised{Day Month Year}
\end{history}

\begin{abstract}
We argue against the rapidly spreading idea of gauge-invariant-extension
(GIE) approach in the nucleon spin decomposition problem, which implies
the existence of infinitely many gauge-invariant
decomposition of the nucleon spin.
\end{abstract}

\ccode{PACS numbers: 12.38.-t, 12.20.-m, 14.20.Dh, 03.50.De}

\section{Introduction}

We now believe that only 1/3 of the nucleon spin comes from the
intrinsic spin of quarks. What carries the remaining 2/3 of the
nucleon spin, then ? This is one of the fundamental questions of QCD.
To answer this question unambiguously, we must first clarify
the following issues. 
What is a precise definition of each term of the decomposition in QCD ?
How can we extract individual term by means of  direct measurements ?
Let us call it the nucleon spin decomposition problem.
Since QCD is a color SU(3) gauge theory, the color gauge-invariance plays a
crucial role in this problem.  
The reason is because the general
gauge-invariance is a necessary condition of observability.
Unfortunately, it is a very delicate problem, which is still under debate.
The conflict appears to lie in the interpretation of the idea
of gauge-invariance.

\section{Nucleon spin decomposition problem and its status}

It has been long known that there are two popular decompositions of the
nucleon spin. One is the Jaffe-Manohar decomposition \cite{ref:JM90},
and the other is the Ji decomposition \cite{ref:Ji97}.
In these two decompositions, only the intrinsic
quark spin part is common and the other parts are all different.
A disadvantage of the Jaffe-Manohar decomposition is that
each term is not separately gauge-invariant except for the
quark spin part. On the other hand, each term of the Ji decomposition is
separately gauge-invariant. Unfortunately, it was claimed and has been
widely believed that
further gauge-invariant decomposition of $J^g$ into its spin and orbital
parts is impossible.

%
%
%
%

Several years ago, however, Chen et al. proposed a new gauge-invariant complete
decomposition of the nucleon spin \cite{ref:Chen08,ref:Chen09}.
The basic idea is a decomposition of the total gluon field into the
physical and pure-gauge components as
$\bm{A} (x) \, = \, \bm{A}_{phys} (x) \, + \, \bm{A}_{pure} (x)$,
which is a sort of generalization of the decomposition of  photon field in QED into the transverse $\bm{A}_\perp$ and longitudinal components $\bm{A}_\parallel$.
%
%
Prominent features of their decomposition is that each term is separately
gauge-invariant. 
It reduces to gauge-variant Jaffe-Manohar decomposition in a particular gauge
$\bm{A}_{pure} \, = \, 0, \, \bm{A} \, = \, \bm{A}_{phys}$.
Soon after, we have shown that the way of gauge-invariant
decomposition of nucleon spin is not necessarily unique, and proposed
another gauge-invariant decomposition \cite{ref:Waka10}. 
%
%
The difference with the Chen decomposition appears in the orbital parts.
The quark OAM part in our decompositin is the same as that of the
Ji decomposition, while the gluon OAM part contain some extra term in addition
to the canonical part. We call this piece the {\it potential angular momenta}
term, because the QED correspondent of this term is the orbital
angular momentum carried by the electromagnetic field or potential,
appearing in the famous Feynman paradox of
electrodynamics \cite{ref:FeynmanBook}.
An arbitrariness of the spin decomposition arises, because this potential
angular momentum term is solely gauge-invariant.
%
%

Furthermore, we found that we can make a seemingly covariant extension of the
above two gauge-invariant decompositions of QCD angular momentum tensor,
which we call the decompositions (I) and (I\hspace{-.1em}I) \cite{ref:Waka11A}.
The word ``seemingly'' is important here, because the decomposition
$A^\mu (x) \, = \, A^\mu_{phys} (x) \, + \, A^\mu_{pure} (x)$,
which is a foundation of the above gauge-invariant
decompositions, is intrinsically non-covariant or frame-dependent,
as we shall see.
Still, this generalization is useful to find relations to high-energy DIS
observables \cite{ref:Waka11B,ref:Waka12}. 
Moreover, we pointed out that the decomposition (I\hspace{-.1em}I) reduces to any ones
of Bashinsky-Jaffe, of Chen et al., and of Jaffe-Manohar,
after an appropriate gauge-fixing in a suitable Lorentz frame.
Since the quark and gluon OAMs in these
decompositions are physically the canonical orbital angular momenta, they may
be called the ``caninical'' family. 
On the other hand, our decomposition (I) is an extension of the Ji decomposition
in the sense that gluon part can also be gauge-invariantly decomposed into the
orbital and intrinsic spin part. The quark OAM part in this decomposition
contains the full covariant derivative, while the gluon part contains
the seemingly covariant generalization of the potential OAM as well.
Our central claim is that these two decompositions (I) and (I\hspace{-.1em}I)
basically exhaust possible gauge-invariant decompositions of the nucleon spin.
However, an opposing claim has rapidly spread in the community
\cite{ref:Ji12,ref:Lorce13A,ref:Lorce13B}.
The claim is that, since the decomposition of the gauge field into its physical
and pure-gauge components is not unique and there are infinitely many such
decompositions, there are in principle infinitely many GI decompositions of
the nucleon spin.

An argument in favor of the second claim was developed by Ji et al. \cite{ref:Ji12}.
According to them, the Chen decomposition is a gauge-invariant extension
(GIE) of the Jaffe-Manohar decomposition based on the Coulomb gauge,
while the Bashinsky-Jaffe decomposition is a GIE of the Jaffe-Manohar
decomposition based on the light-cone gauge.
Because the way of GIE with use of a path-dependent Wilson line is not unique,
there is no need that the two decompositions give the same physical predictions.
One should recognize the oblique nature of the word ``GIE".
In fact, assume that the Chen decomposition and the Bashinsky-Jaffe decompositions
are two physically inequivalent GIEs of the Jaffe-Manohar decomposition.
This immediately raises following questions. 
What is the physical meaning of extended gauge symmetries ?
Are there plural color gauge symmetries in nature ?
Our viewpoint, which we believe is standard, is that
the color gauge symmetry is an intrinsic property of QCD, which is present
from the beginning and in principle there is no need of extending it.
The gauge symmetry is rather freedoms to be eliminated by gauge-fixing
procedures rather than to be obtained by extension.

Another argument in favor of the existence of infinitely many decompositions
of the nucleon spin was developed by Lorc\'{e} \cite{ref:Lorce13A,ref:Lorce13B}.
According to him, the Chen decomposition is a GIE based on the St\"{u}ckelberg trick.
There is a hidden symmetry called the St\"{u}ckelberg symmetry, under which     
the pure-gauge and physical components respectively transform as follows : 
\begin{eqnarray}
 A^{pure}_\mu (x) \ \rightarrow \ A^{pure}_\mu (x)
 \ + \ 
 \frac{i}{g} \,\,U_{pure} (x) \,U_0^{-1} (x) \,
 [\partial_\mu U_0 (x)] \,U^{-1}_{pure} (x), \\
 A^{phys}_\mu (x) \ \rightarrow \ A^{phys}_\mu (x)
 \ - \ 
 \frac{i}{g} \,\,U_{pure} (x) \,U_0^{-1} (x) \,
 [\partial_\mu U_0 (x)] \,U^{-1}_{pure} (x) .
\end{eqnarray}
Since this transformation leaves the total gluon field unchanged, there can be
infinitely many decompositions of $A_\mu (x)$ into physical
and pure-gauge components and consequently infinitely many decompositions
of the nucleon spin.
We claim and in fact showed that, in the QED case, the Chen decomposition
is not a GIE based on the St\"{u}ckelberg trick.  
(See sect. I\hspace{-.1em}I\hspace{-.1em}I of
\refcite{ref:Waka12} and the whole discussion in \refcite{ref:Waka13}.)

\section{Chen decomposition is not a GIE a la St\"{u}ckelberg}

As is well-known, the vector potential of the photon field can be decomposed
into transverse and longitudinal components as
$\bm{A} = \bm{A}_\perp + \bm{A}_\parallel$,
satisfying the divergence-free and irrotational conditions,  
$\nabla \cdot \bm{A}_\perp = 0, \, \nabla \times \bm{A}_\parallel = 0$.
This transverse-longitudinal decomposition is known to be unique, once the
Lorentz frame of reference is fixed \cite{ref:CohenT89}.
Under a general gauge-transformation given as
\begin{eqnarray}
 A^0 (x) &\rightarrow& A^{\prime 0} (x) \ = \ A^0 (x) \ - \ 
 \frac{\partial}{\partial t} \,\,\omega(x), \ \ 
 \bm{A} (x) \ \rightarrow \ \bm{A}^\prime (x) \ = \ 
 \bm{A} (x) \ + \ \nabla \omega (x), \ \ \ \ \ 
\end{eqnarray}
the transverse and longitudinal components transform as follows,
\begin{eqnarray}
 \bm{A}_\perp (x) \ \rightarrow \ \bm{A}^\prime_\perp (x)
 \ = \ \bm{A}_\perp (x), \ \ 
 \bm{A}_\parallel (x) &\rightarrow& \bm{A}^\prime_\parallel (x)
 \ = \ \bm{A}_\parallel (x) \ + \ \nabla \,\omega (x) ,
\end{eqnarray}
which means that the longitudinal component carries unphysical gauge degrees
of freedom, while the transverse part is gauge-invariant. 

Naturally, the longitudinal-transverse decomposition of the 3-vector potential
is Lorentz-frame dependent. (Anyhow, the whole treatment above
is {\it non-covariant}.)
It is true that a vector field that appears transverse in a certain Lorentz frame
is not necessarily transverse in another Lorentz frame.
Nonetheless, the Lorentz-frame dependence of the transverse-longitudinal
decomposition should not make any trouble, because one can start this decomposition
in an arbitrarily chosen Lorentz frame.
After all,  the gauge- and frame-independence of  observables is the core of
the celebrated Maxwell's electrodynamics as a Lorentz-invariant gauge theory.

This QED example indicates that, as long as we are working in a chosen Lorentz frame, there is no arbitrariness in the decomposition of $A^\mu$, as arising from the St\"{u}ckelberg-like transformation of Lorc\'{e} \cite{ref:Lorce13A,ref:Lorce13B}.
In fact, first note that the St\"{u}ckelberg transformation in the abelian case
reduced to the following simple form : 
\begin{eqnarray}
 A^{pure}_\mu (x) \ &\rightarrow& \ A^{pure,g}_\mu (x) \ = \ 
 A^{pure}_\mu (x) \ - \ \partial_\mu C(x), \\
 A^{phys}_\mu (x) \ &\rightarrow& \ A^{phys,g}_\mu (x) \ = \ 
 A^{phys}_\mu (x) \ + \ \partial_\mu C(x) ,
\end{eqnarray}
with $C(x)$ being an arbitrary function of space-time.
This certainly does not change the sum of the physical and
pure-gauge components,
Under this St\"{u}ckelberg, however, the longitudinal and transverse components
transform as follows : 
\begin{eqnarray}
 \bm{A}_\parallel (x) \ &\rightarrow& \ \bm{A}^g_\parallel (x)
 \ \,= \ \bm{A}_\parallel (x) \,\,\ - \ \nabla C(x), \\
 \bm{A}_\perp (x) \ &\rightarrow& \ \bm{A}^g_\perp (x)
 \ = \ \bm{A}_\perp (x) \ + \ \nabla \,C(x) .
\end{eqnarray}
One sees that this transformation leaves the irrotational property of the
longitudinal component unchanged : 
\begin{eqnarray}
 \nabla \times \bm{A}^g_\parallel (x) \ = \ 
 \nabla \times (\bm{A}_\parallel (x) \ - \ \nabla C(x))
 \ = \ \nabla \times \bm{A}_\parallel (x).
\end{eqnarray}
However, the divergence-free or the transversity condition is not
preserved by this transformation,
\begin{eqnarray}
 \nabla \cdot \bm{A}^g_\perp (x) \, &=& \,  
 \nabla \cdot (\bm{A}_\perp (x) \, + \, \nabla \,C(x))
 \, = \, \nabla \cdot \bm{A}_\perp (x) \, + \, 
 \Delta \,C(x) \ \neq \ 
 \nabla \cdot \bm{A}_\perp (x), \ \ \ \ \ \ 
\end{eqnarray}
unless the function $C(x)$ satisfies
the 3-dimensional Laplace equation. This means that we can take $C(x) = 0$
without loss of generality, so that there is no arbitrariness of
St\"{u}ckelberg transformation. 
The fact is that, while the pure-gauge part changes arbitrarily under the gauge-transformation, the physical part is essentially a unique object,
constrained by the transversality condition.

\section{What is needed to settle the controversies}

What is needed to settle the controversies ?
We recall that the main criticism from the GIE approach with use of the
Wilson-line is that the decomposition $A_\mu (x) = A_\mu^{phys} (x) + 
A_\mu^{pure} (x)$ is not unique at all, i.e.
there are infinitely many such decompositions arising from infinitely many
choices of paths.  
From a physical viewpoint, however, the massless gauge field has only two physical or transverse degrees of freedom, and other components are unphysical gauge degrees of freedom.
The standard gauge-fixing procedure is essentially the process of projecting out
the two transverse or physical components of gauge field.
Corresponding to the fact that there exist many gauge-fixing procedures, the
expression of the physical component is not naturally unique. 
Nevertheless, an important wisdom is that final physical predictions for gauge invariant quantities are independent of the choice of gauges ! 

To reveal a hidden problem of the GIE approach, we briefly overview DeWitt's
gauge-invariant formulation of QED \cite{ref:DeWitt62}.
For a given set of electron and photon fields, he constructed a gauge-invariant
set of those in the following manner : 
\begin{eqnarray}
 \psi^\prime (x) \ &\equiv& \ e^{\,i \,\Lambda (x)} \,\psi (x), \ \ 
 A^\prime_\mu (x) \ \equiv \ A_\mu (x) \ + \ \partial_\mu \Lambda (x) ,
\end{eqnarray}
with
\begin{eqnarray}
 \Lambda (x) \ = \ - \,\int_{- \,\infty}^0 \,
 A_\sigma (z) \,\frac{\partial z^\sigma}{\partial \xi} \,
 d \xi ,
\end{eqnarray}
where $z^\mu (x,\xi)$ stands for a path satisfying the following boundary condition : 
\begin{eqnarray}
 z^\mu (x, 0) \ = \ x^\mu, \ \ \ 
 z^\mu (x, - \,\infty) \ = \ \mbox{\rm spatial infinity} .
\end{eqnarray}
The problem is that, while $\psi^\prime (x)$ and
$A_\mu^\prime (x)$ are gauge-invariant by
construction, they are generally {\it path-dependent}.

The path-dependence can easily be understood by considering the simplest case of constant-time paths, which amounts to taking the following $\Lambda (x)$,
\begin{eqnarray}
 \Lambda (x) \ = \ - \,\int_{- \,\infty}^x \,
 \bm{A} (x^0, \bm{z}) \cdot d \bm{z} .
\end{eqnarray}
Let us introduce two GI electron fields corresponding to two different choices
of paths $L_1$ and $L_2$ : 
\begin{eqnarray}
 \psi^\prime (x \,; L_1) &=& \exp \,\left[\,
 - \,i \,e \,\int^x_{L_1} \,\bm{A} (x^0,\bm{z}) \cdot d \bm{z}
 \,\right] \,\psi (x), \\
 \psi^\prime (x \,; L_2) &=& \exp \,\left[\,
 - \,i \,e \,\int^x_{L_2} \,\bm{A} (x^0,\bm{z}) \cdot d \bm{z}
 \,\right] \,\psi (x) .
\end{eqnarray}
The relation between these two electron fields is given by
\begin{eqnarray}
 \psi^\prime (x \,;L_1) \ = \ \exp \,\left[\,
 i \,e \,\left( \int_{L_1}^x \ - \ \int_{L_2}^x \right) \,
 \bm{A} (x^0,\bm{z}) \,\right] \,\psi^\prime (x \,;L_2) .
\end{eqnarray}
Closing the path to a loop $L$ by a connection at spatial infinity,
we get the following relation : 
\begin{eqnarray}
 \psi^\prime (x \,;L_1) &=& \exp \,\left[\,
 i \,e \,\oint_L \,\bm{A} (x^0,\bm{z}) \cdot d \bm{z} \,\right] \,
 \psi^\prime (x \,; L_2) \nonumber \\
 &=& \exp \,\left[\,i \,e \,\int \!\!\int_S \,
 (\nabla_z \times \bm{A} (x^0,\bm{z})) \cdot d \bm{z} \,\right] \,
 \psi^\prime (x \,;L_2) \nonumber \\
 &=& \exp \,\left[\,i \,e \,\int \!\!\int_S \,
 \bm{B} (x^0, \bm{z}) \cdot d \bm{z} \,\right] \,
 \psi^\prime (x \,;L_2) .
\end{eqnarray}
Here, we have used the Stokes theorem.
Since the magnetic flux does not vanish in general, $\psi^\prime (x)$ is
generally path-dependent.
 
Why is the path-dependence a problem ?
This is because the past researches clearly show that the path-dependence
is a reflection of the gauge-dependence 
\cite{ref:Belinfante62,ref:Mandelstam62,ref:RohrlichStrocchi65,ref:Yang85}.
However, there are some nontrivial choices of the function $\Lambda (x)$, which
leads to path-independent set of electron and photon fields,
The first example is given by the choice \cite{ref:KT94}, 
\begin{eqnarray}
 \Lambda (x) \ = \ - \,\int_{- \,\infty}^x \,
 \bm{A}_\parallel (x^0,\bm{z}) \cdot d \bm{z} .
\end{eqnarray}
Here, $\bm{A}_\parallel (x)$ is the the longitudinal component of the photon.
Since the closed-loop line integral of
$\bm{A}_\parallel (x)$ vanishes as
\begin{eqnarray}
 \oint_L \,\bm{A}_\parallel (x^0,\bm{z}) \cdot d \bm{z}
 \ = \ \int \!\! \int_S (\nabla_z \times \bm{A}_\parallel (x^0,\bm{z}))
 \cdot d \bm{S} \ = \ 0 ,
\end{eqnarray}
due to the property $\nabla \times \bm{A}_\parallel = 0$, the electron field $\psi^\prime (x)$
defined with the above $\Lambda (x)$ is not
only gauge-invariant but also
path-independent !

If one remembers the familiar formula for the longitudinal component,
\begin{eqnarray}
 \bm{A}_\parallel (x) \ = \ \nabla \,\,\frac{1}{\nabla^2} 
 \,\nabla \cdot \bm{A} (x), \ \ \ 
 \bm{A}_\perp (x) \ = \ \bm{A} (x) \ - \ \bm{A}_\parallel (x) ,
\end{eqnarray}
one can also express as follows : 
\begin{eqnarray}
 \psi^\prime (x) &=& \exp \,\left[\, - \,e \,
 \int_{- \,\infty}^x \,\left( \nabla_z \,
 \frac{1}{\nabla_z^2} \,\nabla_z \cdot \bm{A} (x^0,\bm{z})
 \right) \cdot d \bm{z} \,\right] \,\psi (x) \nonumber \\ 
 &=& \exp \,\left[\, - \,i \,e \,\,
 \frac{\nabla \cdot \bm{A}}{\nabla^2} \,(x) \right] \,\psi (x) .
\end{eqnarray}
In this form, the path-independence of the
GI electron field is self-evident. Note that this $\psi^\prime (x)$ is
nothing but the GI {\it physical electron} introduced by Dirac 
\cite{ref:Dirac58,ref:LavelleMcMullan93}.
Using the same function $\Lambda (x)$, the GI potential $A^\prime_\mu (x)$
becomes 
\begin{eqnarray}
 \bm{A}^\prime (x) 
 \ = \ \bm{A}_\perp (x), \ \ \  
 A^{\prime 0} (x) \ = \ A^0 (x) \ + \ \int_{- \,\infty}^x \,
 \bm{A}_\parallel (x^0, \bm{z}) \cdot d \bm{z} .
\end{eqnarray}
One thus reconfirms that the physical component of the spatial part of the photon field
is nothing but the familiar transverse component.

Also interesting is the following second example.
Using a constant 4-vector $n^\mu$, we introduce the decomposition : 
\begin{eqnarray}
 A_\mu (x) \ = \ A^{phys}_\mu (x) \ + \ A^{pure}_\mu (x)
 \ \equiv \ (\,P_{\mu \nu} \ + \ Q_{\mu \nu} \,) \,A^\nu (x) ,
\end{eqnarray}
with
\begin{eqnarray}
 P_{\mu \nu} \ = \ g_{\mu \nu} \ - \ 
 \frac{\partial_\mu \,n_\nu}{n \cdot \partial}, \ \ \ 
 Q_{\mu \nu} \ = \ \frac{\partial_\mu \,n_\nu}{n \cdot \partial}.
\end{eqnarray}
These two components satisfy the important properties : 
\begin{eqnarray}
 n^\mu \,A_\mu^{phys} (x) \ = \ 0 ,
 \ \ \ \ 
 \partial_\mu \,A_\nu^{pure} (x) \ - \ 
 \partial_\nu \,A_\mu^{pure} \ = \ 0 . 
\end{eqnarray}
Now, we propose to take the following $\Lambda (x)$,
\begin{eqnarray}
 \Lambda (x) \ = \ - \,\int_{- \,\infty}^x \,
 A_\mu^{pure} (z) \,d z^\mu ,
\end{eqnarray}
and define the GI electron and photon fields by Eq.(11). 
Note that, by using the Stokes theorem in 4 space-time dimension,
the following identity holds
\begin{eqnarray}
 \oint_L \,A_\mu^{pure} (z) \,d z^\mu \ = \ 
 \frac{1}{2} \int \!\! \int_S \,\left(
 \partial_\mu \,A_\nu^{pure} \ - \ \partial_\nu \,A_\mu^{pure}
 \right) \,d \sigma^{\mu \nu} \ = \ 0 ,
\end{eqnarray}
so that this $\Lambda (x)$ turns out to be
path-independent. In fact, $\Lambda (x)$ can also be expressed in the
following form : 
\begin{eqnarray}
 \Lambda (x) &=& - \,\int_{- \,\infty}^x \,
 \frac{\partial^z_\mu \,n_\nu}{n \cdot \partial^z} \,
 A^\nu (z) \,d z^\mu \ = \ 
 \frac{n \cdot A(x)}{n \cdot \partial} .
\end{eqnarray}
The GI electron and photon fields are then given by
\begin{eqnarray}
 \psi^\prime (x) &=& 
 e^{i \,e \,\frac{n \cdot A(x)}{n \cdot \partial}} \,\psi (x), \ \ \ 
 A_\mu^\prime (x) \ = \ \left( g_{\mu \nu} \ - \ 
 \frac{\partial_\mu \,n_\nu}{n \cdot \partial} \right) \,A^\nu (x)
 \ = \ A_\mu^{phys} (x). \ \ \ \ 
\end{eqnarray}
Note that the physical component satisfies the following condition,
\begin{equation}
 n^\mu \,A_\mu^{phys} (x) \ = \ 0 ,
\end{equation}
which is nothing but the gauge-fixing condition in general axial gauge.

These two examples clearly show that the form of the physical component is not in
fact unique. It is expressed in several different forms, which is not unrelated
to the fact that there are many gauge-fixing procedures in different Lorentz frame. 
Nevertheless,  standard belief  is that, as far as we handle the gauge- and Lorentz-invariant quantity in a usual sense,  the final prediction should be
the same.
After this pedagogical introduction, we now want to address our
central question.
Is the gluon spin term appearing in the longitudinal nucleon spin sum rule such a quantity with standard gauge-invariance or not ?
To answer this question, we must generalize the construction of $A^{phys}_\mu (x)$
to the nonabelian gauge theory.   
We point out that, in the past, tremendous efforts have been made to figure out the two physical components of the gauge field.

Especially useful for our purpose is the geometrical construction
by Ivanov, Korchemsky, and Radyushkin based on the fiber-bundle formulation
of gauge theories \cite{ref:IKR86}.
In their formulation, the gauge-covariant gluon field can be constructed in the
following form : 
\begin{eqnarray}
 A^g_\mu (x) &=& A_\nu (x_0) \,\frac{\partial x^\nu_0}{\partial x^\mu}
 \ - \ \int_{x_0}^x \,d z^\nu \,\frac{\partial z^\rho}{\partial x^\mu} \,
 W_C (x_0, z) \,F_{\nu \rho} (z \,; A) \,W_C (z, x_0) ,
\end{eqnarray}
where
\begin{eqnarray}
 W_C (x,x_0) \ \equiv \ P \,\exp \,\left[\,
 i \,g \,\int_{x_0}^x \,d z^\mu \,A_\mu (z) \,\right] ,
\end{eqnarray}
is a familiar Wilson line with $z (s)$
being a path $C$ in 4-dimensional space-time with an appropriate boundary condition.
One should clearly keep in mind the fact that $A^g_\mu (x)$ so constructed is generally dependent of the choice of path $C$.
However, these authors clearly recognize the fact that the choice of suitable
path in the geometrical formulation corresponds to gauge-fixing procedure.
They also showed that, with some natural choices of paths, the above way of fixing
the gauge is equivalent to taking gauges satisfying a particularly simple
condition $W_C (x, x_0) = 1$.
This class of gauge is called the {\it contour gauge} and it is shown to have an attractive feature that they are {\it ghost-free}.
Some familiar gauges belonging to the contour gauge are the Fock-Schwinger
gauge, the Hamilton gauge, and the axial gauge.
In particular, the axial gauge corresponds to taking an infinitely long
straight-line path $z^\mu (s) \, = \, x^\mu \ + \ s \,  n^\mu \ (0 < s < \infty)$.
This gives the following expression for $A^g_\mu (x)$ : 
\begin{eqnarray}
 A^g_\mu (x) \ = \ n^\nu \,\int_0^\infty \,
 W_C^\dagger (x + n \,s, \infty) \,
 F_{\mu \nu} (x + n \,s \,; A) \,
 W_C (x + n \,s, \infty) ,
\end{eqnarray}
with
\begin{eqnarray}
 W_C (x, \infty) \ = \ P \,\exp \,\left(
 i \,g \,\int_0^\infty \,d s \,\,n^\mu \,A_\mu ( x + n \,s)
 \right) .
\end{eqnarray}
Using the antisymmetry of the field-strength tensor, it is easy to verify
the identity $n^\mu \,A^g_\mu = 0$, which is 
nothing but the gauge-fixing condition in {\it general axial gauge}.
Since $n^\mu$ is an arbitrary constant 4-vector, it contains several popular
gauges, i.e. the temporal gauge, the light-cone gauge, and the spatial axial-gauge, respectively corresponding to the
choice $n^\mu \, = \, (1,0,0,0), n^\mu \, = \, (1,0,0,1) / \sqrt{2}$ and
$n^\mu \, = \, (0,0,0,1)$.

Since our main interest here is to show the traditional gauge-invariance of the evolution equation of the longitudinal gluon spin, let us inspect the perturbative  (lowest order) contents of the defining equation of the physical component
$A^{phys}_\mu (x) \equiv A^g_\mu (x)$, which reduces to
\begin{eqnarray}
 A^{phys}_\mu (x) \simeq \ n^\nu \int_0^\infty \,d s \,
 \left( \partial_\mu A_\nu (x + n \,s) \ - \ 
 \partial_\nu \,A_\mu (x + n \,s) \right) .
\end{eqnarray}
Introducing the Fourier transform, this physical component can
be expressed as, 
\begin{eqnarray}
 A^{phys}_\mu (x) 
 &\simeq& n^\nu \,\int_0^\infty \,d s
 \int \,\frac{d^4 k}{(2 \,\pi)^4} \,
 e^{\,i \,k \cdot (x + n \,s)} \,\left(
 i \,k_\mu \,\tilde{A}_\nu (k) \ - \ 
 i \,k_\nu \,\tilde{A}_\mu (k) \right) \nonumber \\
 &=& \int \,\frac{d^4 k}{(2 \,\pi)^4} \,
 \left( g_{\mu \nu} \ - \ \frac{k_\mu \,n_\nu}{k \cdot n}
 \right) \,\tilde{A}^\nu (k) 
 \ = \ 
 \left( g_{\mu \nu} \ - \ 
 \frac{\partial_\mu \,n_\nu}{n \cdot \partial} \right) \,
 A^\nu (x) . \ \ \ \ \ 
\end{eqnarray}
(Note that, this is an exact expression in the abelian
case.)
This in turn gives the lowest order expression for the physical gluon
propagator as follows,
\begin{eqnarray}
 \langle T \,(A^{phys}_{\mu,a} (x) \,
 A^{phys}_{\nu,b} (y) ) \rangle^{(0)} \ = \ 
 \int \,\frac{d^4 k}{(2 \,\pi)^4} \,
 e^{\,i \,k \,(x - y)} \,
 \frac{- \,i \delta_{a b}}{k^2 + i \,\varepsilon}
 \,P_{\mu \nu} (k) ,
\end{eqnarray}
with
\begin{eqnarray}
 P_{\mu \nu} (k) \ = \ g_{\mu \nu} \ - \ 
 \frac{k_\mu \,n_\nu + n_\mu \,k_\nu}{k \cdot n}
 \ + \ \frac{n^2 \,k_\mu \,k_\nu}{(k \cdot n)^2} ,
\end{eqnarray}
which is nothing but the gluon propagator in the general
axial gauge.
In this way, the {\it path dependence} or {\it direction dependence}
in the geometric formulation is replaced by the {\it gauge dependence}
within the general axial gauge.
In this setting, the gluon spin operator reduces to the form : 
\begin{eqnarray}
 M^{\lambda \mu \nu}_{G-spin} \ = \ 2 \,\mbox{Tr} \,
 \left[ F^{\lambda \nu} \,A^\mu \ - \ F^{\lambda \mu} \,A^\nu
 \right] .
\end{eqnarray}
In this equation, $A^\mu$ should be regarded as the physical gluon field
satisfying the general axial gauge condition.

\section{Evolution equation for the gluon spin in general axial gauge}

Now, we are ready to investigate the evolution equation for the quark and gluon
spins in general axial gauge. Let us start with the following covariant relation,
\begin{eqnarray}
 \langle P s \,|\, M^{\lambda \mu \nu} (0) \,|\, P s \rangle
 \ = \ J_N \,\frac{P_\rho \,s_\sigma}{M_N^2} \,
 \left[ \,2 \,P^\lambda \,\epsilon^{\nu \mu \rho \sigma}
 \ - \ P^\mu \,\epsilon^{\lambda \nu \rho \sigma} \ - \ 
 P^\nu \,\epsilon^{\mu \lambda \rho \sigma} \,\right] .
\end{eqnarray}
The longitudinal nucleon spin sum rule is obtained by setting $\mu = 1, \nu = 2$,
and by contracting with the constant 4-vector $n^\mu$, which gives
\begin{eqnarray}
 J_N \ = \ \frac{1}{2} \ = \ 
 \frac{\langle P s \,|\,n_\lambda \,M^{\lambda 1 2} (0) \,
 |\,P s \rangle}{2 \,{P \cdot n}} .
\end{eqnarray}
An important fact is that this last equation is no longer a covariant relation.
The quantity $n_\lambda$ appearing in this equation should be identified with
the 4-vector  that characterizes the Lorentz-frame, in which the gauge-fixing condition $n^\mu \,A_\mu = 0$ is imposed.
In this setting, we have calculated the 1-loop anomalous dimension of the above
gluon spin operator, and found that it reproduces the commonly-known answer,
irrespectively of the choice of $n^\mu$.
Although this is a proof within a restricted class of gauge, i.e. the general axial gauge, characterized by a constant 4-vector $n^\mu$,  it strongly indicates that the gluon spin term in the longitudinal nucleon spin sum rule is a gauge-invariant quantity in a true or traditional sense.
This is a welcome conclusion, because it means that now there is no conceptual
conflict between the observability of the nucleon spin decomposition (I)
and the general gauge principle.

\section{Conclusion}

We have carried out a detailed comparison of the two fundamentally different approaches to the nucleon spin decomposition problems, i.e. the GIE approach
and the standard gauge-fixing approach.
If both give the same answer, there is no practical problem. However, if they give different answers, one must stop and think it over.
In our opinion, conceptually legitimate is the latter approach. For, there is only one
color gauge symmetry of QCD, which is present from the beginning.
This gauge symmetry is rather freedoms to be eliminated by gauge-fixing procedures rather than to be gained by extension.
This general consideration gives a support to our claim that there are only two (not infinitely many) physically inequivalent GI decompositions (I) and
(I\hspace{-.1em}I) of the nucleon spin.

\section*{Acknowledgments}
The author would like to greatly appreciate many stimulating discussions
with Cedric Lorc\'{e}, and also with Takahiro Kubota.


\end{document}